\documentclass[aps,prb,showpacs,preprintnumbers,amsmath,amssymb,twocolumn,letterpaper]{revtex4}
\usepackage{amssymb}
\usepackage{ifpdf}
\usepackage[utf8]{inputenc}
\usepackage{graphicx}
\usepackage{hyperref}
\usepackage{bm}
\usepackage{graphicx,color}

\begin{document}

\newcommand{\OLD}[1]{{\tiny {\bf old:} #1 }}
\newcommand{\NEW}[1]{{ \it #1 }}
\renewcommand{\vec}[1]{{\bf #1}}
\newcommand{\w}{\omega}
\newcommand{\Tc}{$T_{c}$ }
\newcommand{\rhoxx}{$\rho_{xx}$ }
\newcommand{\rhoxy}{$\rho_{xy}$ }

\renewcommand{\floatpagefraction}{0.5}

\title{Helimagnon Bands as Universal Excitations of Chiral Magnets}

\author{M. Janoschek}
\affiliation{Physik Department E21, Technische Universit\"at M\"unchen, D-85748 Garching, Germany}
\affiliation{Laboratory for Neutron Scattering, Paul Scherrer Institut \& ETH Zurich, CH-5232, Villigen, PSI}
\altaffiliation[Present address: ]{Department of Physics, University of California, San Diego, La Jolla, CA 92093-0354, USA; mjanoschek@physics.ucsd.edu}

\author{F. Bernlochner}
\affiliation{Physik Department E21, Technische Universit\"at M\"unchen, D-85748 Garching, Germany}

\author{S. Dunsiger}
\affiliation{Physik Department E21, Technische Universit\"at M\"unchen, D-85748 Garching, Germany}

\author{C. Pfleiderer}
\affiliation{Physik Department E21, Technische Universit\"at M\"unchen, D-85748 Garching, Germany}

\author{P. B\"oni}
\affiliation{Physik Department E21, Technische Universit\"at M\"unchen, D-85748 Garching, Germany}

\author{B. Roessli}
\affiliation{Laboratory for Neutron Scattering, Paul Scherrer Institut \& ETH Zurich, CH-5232, Villigen, Switzerland}

\author{P. Link}
\affiliation{Forschungsneutronenquelle Heinz Maier-Leibniz (FRM II), Technische Universit\"at M\"unchen, D-85748
Garching, Germany}

\author{A. Rosch}
\affiliation{Institute for Theoretical Physics, Universit\"at zu K\"oln, Germany\\
Kavli Institute for Theoretical Physics, Santa Barbara, USA}

\date{\today}

\begin{abstract}
MnSi is a cubic compound with small magnetic anisotropy, which stabilizes a helimagnetic spin spiral that reduces to a ferromagnetic and antiferromagnetic state in the long- and short-wavelength limit, respectively. We report a comprehensive inelastic neutron scattering study of the collective magnetic excitations in the helimagnetic state of MnSi. In our study we observe a rich variety of seemingly anomalous excitation spectra, as measured in well over twenty different locations in reciprocal space. Using a model based on only three parameters, namely the measured pitch of the helix, the measured ferromagnetic spin wave stiffness and the amplitude of the signal, as the only free variable, we can simultaneously account for \textit{all} of the measured spectra in excellent quantitative agreement with experiment. Our study identifies the formation of intense, strongly coupled bands of helimagnons as a universal characteristic of systems with weak chiral interactions.
\end{abstract}


\vskip2pc

\maketitle

\section{Introduction}
\begin{figure}[h!]
\centering
\includegraphics[width=.47\textwidth,clip=]{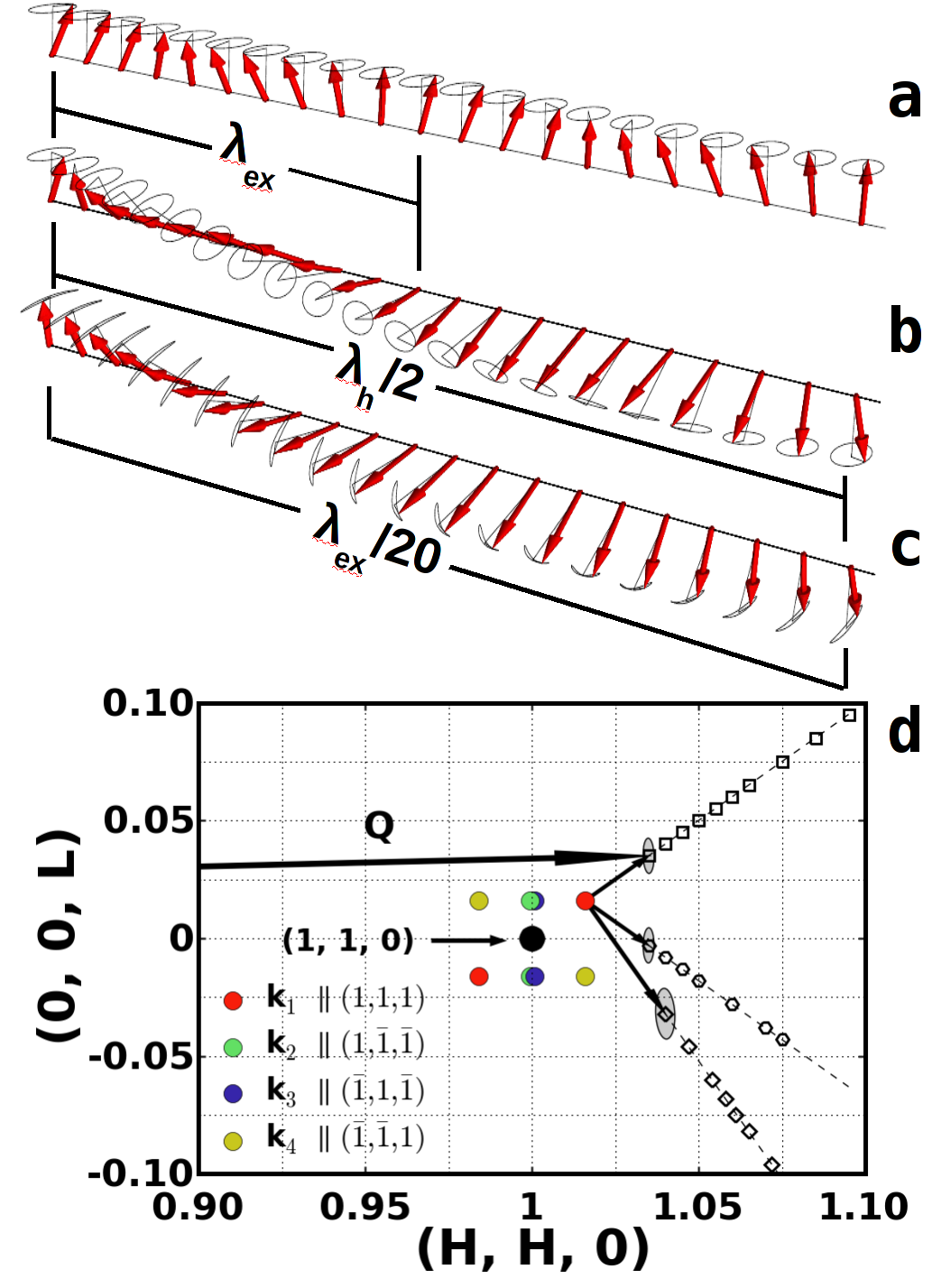}
\caption{ (a) Depiction of a low-energy spin wave excitation of a ferromagnet with wavelength $\lambda_{\rm ex}$. The spins precess around the uniform magnetization. (b+c) Sketch of a similar excitation in the case of a helimagnet with a momentum, ${\rm q}$, parallel to the propagation vector of the helix,  ${\rm k_h}$. The wavelength $\lambda_{\rm ex}$ is smaller (b) or much larger (c) than the pitch of the helix $\lambda_{\rm h} = \frac{2\pi}{k_{\rm h}}$. Note that only for this parallel configuration the excitations of the ferromagnet and the helimagnet are very similar for short $\lambda_{\rm ex}$. (d) Reciprocal space map around the nuclear $(110)$ Bragg peak. The locations of the magnetic Bragg satellites $\bf{k}_1$ to $\bf{k}_4$ associated with the four domain configurations of the helimagnetic order are shown as color coded throughout the text. The locations at which the spin excitations were measured are shown by the open symbols. They may be grouped along three trajectories, where the resolution ellipsoids are shown in gray shading. Note that \textit{all of the spectra recorded} may be simultaneously accounted for by the model described in the text using just one parameter (the intensity).}\label{figure1}
\end{figure}
The spontaneous breaking of a continuous symmetry in a magnetically ordered state implies the existence of Goldstone modes. Thus, the low-energy spin wave dispersions of ferromagnets and commensurate antiferromagnets are universal -- they can be deduced from simple symmetry arguments and the description does not depend on microscopic details \cite{bloch:30,holstein:40,dyson:56,brockhouse:57,okazaki:64}. In ferromagnets, where the order parameter is a conserved quantity, the resulting spin waves have a quadratic dispersion while in antiferromagnets the superposition of the normal modes of the sublattices leads to the well-known linear dispersion at low energies \cite{kittel,keffer:53}.\\
In recent years complex forms of magnetic order related to weak chiral interactions of non-centrosymmetric systems, also referred to as Dzyaloshinsky-Moriya (DM) interaction attract great interest. Systems in which DM interactions play an important role include multiferroics \cite{Yamasaki:07}, the parent compounds of the high-temperature superconductors \cite{Juricic:04}, thin magnetic films (Mn on a W substrate, or Fe on a Ir substrate \cite{Bode:07,Bergmann:06}), heavy fermion superconductors \cite{Pfleiderer:09} and even itinerant-electron magnets, like MnSi, which displays a skyrmion lattice at ambient pressure \cite{muehlbauer:09} and an extended non-Fermi liquid resistivity at high pressures\cite{pfleiderer:01,Doiron:03}. As DM interactions are of growing importance in non-centrosymmetric systems this raises the question for any universal properties of the spin excitations in systems with weak chiral interactions.\\
Helimagnetic order is very well established in systems with magnetic frustration such as rare earth elements like Tb or Ho. An excellent experimental and theoretical account has been given by Jensen and Mackintosh \cite{jensen:91,moller:67}, but excitation spectra perpendicular to the magnetic propagation vector have not been studied before. In addition it has long been appreciated that helimagnetic order due to frustration differs fundamentally from helimagnetic order due to DM interactions.\cite{kataoka:87} The work on frustrated magnets has been contrasted by studies of the tetragonal system Ba$_2$CuGe$_2$O$_7$ \cite{Zheludev:99}, in which the influence of DM interactions on the excitation spectra has been addressed. However, Ba$_2$CuGe$_2$O$_7$ orders antiferromagnetically on local length scales. Due to the much stiffer spectrum of an antiferromagnet compared to a ferromagnet, the effects of DM interactions in a helical antiferromagnet are therefore substantially weaker and qualitatively different as compared to a helical ferromagnet like MnSi. Moreover, the helimagnetic order in Ba$_2$CuGe$_2$O$_7$ is distorted and shows higher order harmonic contributions of at least 20\,\%.\\
We have therefore decided to study the perhaps simplest example for the effects of DM interactions, notably cubic chiral helimagnets, where the uniform magnetization rotates slowly around an axis with a characteristic wave vector $\vec{k}_h$. As illustrated in Fig.\,\ref{figure1}, a helical state is essentially a ferromagnet on short length scales and an antiferromagnet for long distances. In fact, from a more general viewpoint, all forms of complex order can be interpreted as a superposition of helimagnetic order \cite{sandratskii:98}. In cases where the pitch of the helix is much larger than the lattice spacing, $\lambda_h\gg a$, a helix can become remarkably stable against crystalline anisotropies. The Goldstone theorem ensures that the excitations of an incommensurate helimagnet are gapless (in case the helimagnet is commensurate the gap is exponentially small) \cite{belitz:06,commentM}. In this paper we report, that the spin excitations for a helix with a long pitch display an universal property: the formation of strongly coupled bands of helimagnons.\\
The B20 compound MnSi is ideally suited to study the collective spin excitations of helimagnets due to DM interactions experimentally, because the magnetic properties result from a clear separation of energy scales in a metallic host. MnSi crystallizes in the non-centrosymmetric cubic space group P2$_1$3 ($a=4.558\,{\rm \AA}$). Below $T_c=29.5\,{\rm K}$ and in zero magnetic field a long-wavelength spin spiral with the spins perpendicular to the propagation direction stabilizes. The periodicity $\lambda_h\approx180\,{\rm \AA}$ of the helix results from the competition of ferromagnetic exchange interactions, as the strongest scale, and Dzyaloshinskii-Moriya (DM) interactions as a manifestation of weak spin- orbit coupling in crystal structures without inversion center, on an intermediate scale  \cite{Shirane:Dec83,Ishida:85}. The propagation direction of the spin spiral is finally locked to the cubic space diagonal through very small crystal field interactions, providing the weakest scale. In comparison with the helical modulation, the Fermi wave-vector is large $\vec{k}_{\rm F}\sim0.7\,{\rm \AA}^{-1}$.\\
MnSi has recently attracted great interest as the perhaps best candidate displaying a genuine non-Fermi liquid metallic state in a three-dimensional metal at high pressure \cite{pfleiderer:01,Doiron:03,pfleiderer:07}. Moreover, based on neutron scattering and $\mu$-SR at high pressures \cite{pfleiderer:04,uemura:07}, it has been suggested that the NFL behavior may be related to spin textures with non-trivial topology  \cite{binz:06,tewari:06,roessler:06,fischer:08}. In fact, a skyrmion lattice, was recently identified unambiguously at ambient pressure in a small phase pocket just below $T_c$, which is believed to be stabilized by thermal fluctuations \cite{muehlbauer:09}. To resolve the origin of these exciting properties it has become of great importance in its own right to establish at first a full account of the normal helimagnetic state and its spectrum of excitations. This has also inspired the experimental study presented here.\\
The three different energy scales governing the physics of the helimagnetic state of MnSi result in four different regimes for spin excitations. As a function of increasing momentum of the excitations these regimes may be summarized as follows. For the smallest momenta one explores the antiferromagnetic limit, $q \ll k_p$, where a linear spectrum of spin waves is expected.  Here $\vec q$ is the wavevector measured from the ordering vector $\vec{k}_{\rm h}$ of the helix and $k_p$ measures the weak pinning of the orientation of the helix to the crystalline lattice, which is fourth order in spin orbit coupling ($k_p$ is expected to be of the order of $k_h^2/k_F$ and therefore extremely tiny). The antiferromagnetic limit is followed by the helimagnetic regime for $k_p \ll q \ll k_{\rm h}$, where the pinning of the helix can be neglected. Here the universal properties are predicted to consist of a dispersion similar to an antiferromagnet, $\omega\propto\vert \vec{q}\vert^n$ with $n=1$ for propagation parallel to $\vec{k}_{\rm h}$, whereas in the transverse direction the spectrum is quadratic with $n=2$ characteristic of a ferromagnet \cite{belitz:06,maleyev:06}. Further increasing the momenta leads to the cross-over between the helimagnetic regime and ferromagnetic limit, $k_{\rm h}\lesssim q \ll k_{\rm c}$ with $k_c \sim k_F$ deep in the magnetic phase. The magnetic structure is now probed on a length scale short compared to its pitch and a spectrum reminiscent of a ferromagnet may be expected, because the system locally appears to be a ferromagnetically aligned. However, as shown below this expectation differs dramatically from the observed behavior. Instead of ferromagnetic magnons multiple helimagnon bands are excited, which may be regarded as a new universal property of magnetic materials. Finally, for the largest momenta, $q \gtrsim  k_{\rm c}$, the ferromagnetic spin waves cross over into the Stoner continuum as established experimentally in a large number of systems. The properties depend now on details of the Fermi surface, i.e., they are no longer universal.\\
Several pioneering inelastic neutron scattering studies have been carried out in MnSi. In the mid 1970ies Ishikawa and coworkers established for the first time in any magnetic metal the existence of paramagnon fluctuations over major portions of the Brillouin zone \cite{Ishikawa:77}. They further observed ferromagnetic spin waves when suppressing the helimagnetic state in an applied magnetic field \cite{Ishikawa:77}. More recently polarized inelastic neutron scattering established that the spin fluctuations in the paramagnetic state are chiral \cite{roessli:02} and the existence of spin-flip excitations in a small applied field \cite{semadeni:99,roessli:04}.\\ 
The magnetic phase diagram of MnSi is remarkably rich. For small magnetic fields between of 0.05\,T and 0.15\,T a reorientation transition takes place and the helimagnetic modulation aligns with the applied field to form the so-called conical phase. The helical modulation is suppressed for fields above $\sim0.6\,{\rm T}$. Finally, a small phase pocket just below $T_c$ has been identified recently as the first example of a skyrmion lattice, thereby demonstrating the inherent instability of the helical order to stabilize non-trivial spin textures in small magnetic fields.\\
For a comprehensive understanding it is consequently important to measure the spin excitations in the prestine (zero-field cooled) helical state. This has not been attempted before, since four equally populated domains form, all of which contribute to the full spectrum of the spin excitations, as described below. In the following we denote these domains as follows: domain 1 corresponds to $\bf{k}_1\parallel (1,1,1)$, domain 2 corresponds to $\bf{k}_2\parallel (1, \bar{1},\bar{1})$, domain 3 corresponds to $\bf{k}_3\parallel (\bar{1}, 1,\bar{1})$, and domain 4 corresponds to $\bf{k}_4\parallel (\bar{1},\bar{1},1)$ (cf. Fig.~\ref{figure1}). We confirmed that the domains are equally populated in our measurements.\\
The experimental study of the excitation spectra of the helimagnetic state are also extremely challenging because of the long wavelength of the helical modulation. This makes high-resolution measurements in a relatively small portion of reciprocal space around a magnetic satellite reflection defined by $q\simeq k=0.035~{\rm \AA^{-1}}$ necessary. In the study described here we focus on the cross-over between the ferromagnetic limit and the helimagnetic regime described above. In fact, the excitations for even smaller momenta deep in the helimagnetic regime and the antiferromagnetic limit described above are technically not accessible in present-day inelastic neutron scattering measurements, even for the most advanced neutron scattering techniques.
\section{Experimental methods}
\begin{figure*}[!ht]
\centering
\includegraphics[width=\textwidth,clip=]{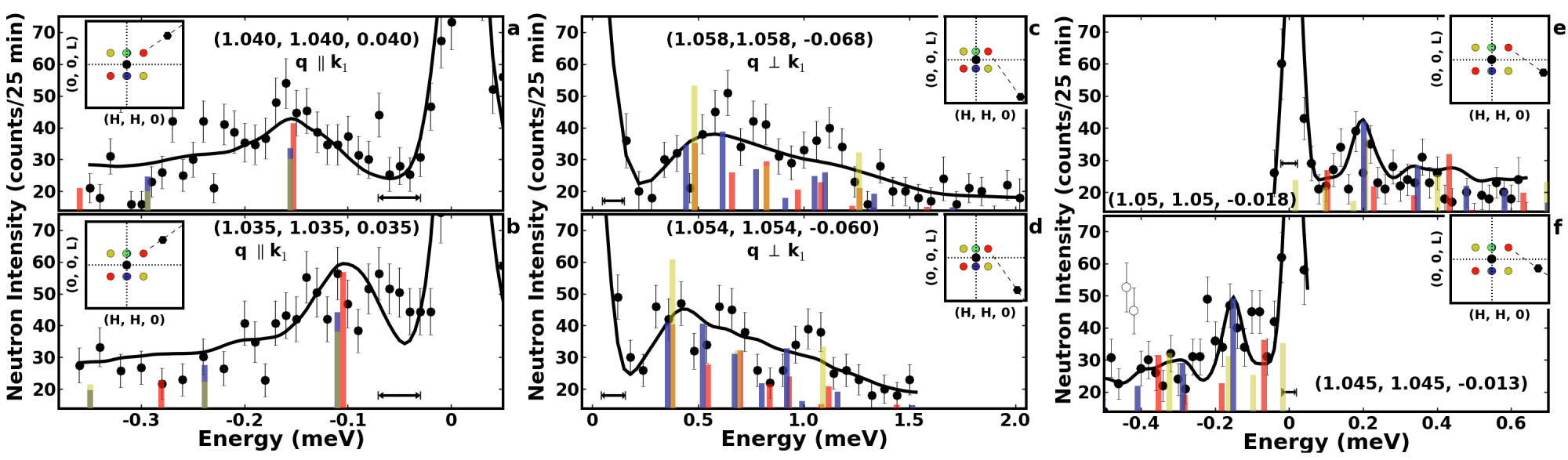}
\caption{Typical constant-$\bm{Q}$ scans at selected locations of the three trajectories shown in
Fig.\,\ref{figure1}. Note that the energy and momentum resolution is at the technical limit currently available.  The strong elastic peak at 0~meV is due to incoherent scattering. The inset in each panel shows the precise location in reciprocal space as a black spot where data were recorded. The curves represent the intensity calculated in the model described in the text, where all data are accounted for by the same values of the ferromagnetic spin wave stiffness, the helical wavelength and the intensity. (a), (b): Data for the trajectory with $\vec{q}\parallel\vec{k}_1$; these scans are dominated by a very broad maximum. (c), (d): Data for $\vec{q}\perp\vec{k}_1$; these scans show very broad, essentially featureless intensity that decreases for increasing energy. (e), (f): Data for an arbitrary trajectory emanating from $\vec{k}_1$; data for this trajectory are characterized by almost featureless intensity over an extremely wide range of energies and a distinct peak in a small range. The open data points below -0.4 meV in (f) represent a spurious signal that arises from additional incoherent scattering of neutrons from the monochromator crystals of the triple axis spectrometer.}\label{figure2}
\end{figure*}
A large single crystal of $\sim8\,{\rm cm^3}$ grown by the Bridgman method was studied. The paramagnetic spin fluctuations of the same single crystal were reported previously in Ref.\,\cite{roessli:02}. The single crystal displayed a lattice mosaic of the order $0.5^{\circ}$. The specific heat, susceptibility and resistivity of small pieces taken from this large single crystal were in excellent agreement with the literature, where the residual resistivity ratio (RRR) was of the order 100. The latter indicates good, though not excellent sample purity.\\
The bulk of our studies was carried out on the cold triple-axis spectrometer TASP at the Paul Scherrer Institut (PSI) \cite{semadeni:01}, with a few supplementary measurements on PANDA at the Forschungsneutronenquelle Heinz-Maier Leibnitz (FRM II). For the experiment the single crystal was cooled with an ILL-type orange He-cryostat. The sample was oriented with the [1,1,0] and [0,0,1] crystallographic directions in the scattering plane. In order to avoid second order contamination of the neutron beam a beryllium filter was inserted between the sample and the analyzer.\\
The experiments were performed around the nuclear Bragg reflection $(1, 1, 0)$ taking advantage of the large magnetic structure factor of the corresponding magnetic Bragg reflections (cf. Ref.~\onlinecite{Ishikawa:77} and Fig.~\ref{figure1}(c)). Two different setups were used. For setup I the triple-axis spectrometer was operated with fixed final wave vector $k_f$ = 1.2~\AA$^{-1}$. Additionally 20' Soller collimators were installed in the incident beam and in front of the analyser whereas 40' collimators were used in front of the detector. In setup II we used $k_f$ = 1.4~\AA$^{-1}$ and a collimation setting: open-40'-MnSi-40'-80'. The energy resolution in setup I and II was approximately 40~$\mu$eV and 90~$\mu$eV, respectively. The corresponding $\bm{Q}$-resolution is depicted in Fig.~\ref{figure1}. It is important to emphasize that the resolution achieved was about the best currently available.\\
In order to compare the model with the experimental results the theoretical scattering function $S(\vec{Q},\omega)$ was convoluted with the resolution function of the triple-axis spectrometers we used\cite{Cooper:67,popovici:75}. For the fits presented in this article a program called TASRESFIT \cite{Janoschek:08} has been developed in PYTHON to perform the convolution of the spectrometer resolution with the theoretical cross-section by means of Monte-Carlo integration.\\ 
Further, we have introduced a Gaussian profile centred at $\hbar\omega = 0.01\,{\rm meV}$ to describe the incoherent scattering together with a constant background for the fit of all data presented. The background was carefully measured at positions in reciprocal space that were not contaminated by inelastic scattering around the $(1, 1, 0)$ Bragg reflection and was estimated as 20 Counts/25 min.\\
Finally, some of the energy scans measured during the course of our experiments were contaminated by spurious scattering that was effectively caused by neutrons that were scattered incoherently from the monochromator crystals of the used triple-axis spectrometer (see e.g. open symbols Fig.~\ref{figure2}(f) at $E\approx -0.4$\,meV). When the spectrometer is set to a non-zero energy transfer, i.e. $\vert \vec{k}_i\vert \neq \vert \vec{k}_f\vert$, but the angle between $\vec{k}_i$ and $\vec{k}_f$ as well as the orientation of the sample correspond to a geometry that allows for Bragg scattering from the sample, the incoherently scattered neutrons from the monochromator that have a wave vector $\vert\vec{k}_{i,inc}\vert = \vert \vec{k}_f\vert$ lead to the observation of accidental Bragg scattering. This type of spurious scattering is well-known as Currat-Axe peaks in the literature \cite{Shirane:02}. For the analysis we considered only scans where the spurious scattering could be separated well from the inelastic signal.
\section{Experimental results}
All scans shown in the following were performed at a temperature of 20~K. This way the intensity was strong while still being significantly below $T_c$ deep inside the helical phase. Summarized in Fig.\,\ref{figure1}(d) are the locations in reciprocal space where we have recorded excitation spectra. A total of 24 spectra were recorded. To cover the regime of interest near the magnetic ordering wave vector we have mostly carried out inelastic scans at reciprocal lattice positions along three traces in reciprocal space as shown in Fig.~\ref{figure1}(d). Two traces (square and rhombus symbols) were selected such that they are parallel and perpendicular to $\bm{k}_1$, while the third trace (circle symbols) was recorded along an arbitrary direction. The latter may be seen as the most stringent test of the theoretical model used to describe the experimental data. All spectra were recorded in energy scans at fixed momentum $\bm{Q}$. The scattering intensity observed experimentally is hence the result of contributions from each of the domains, that is, most of the scattering wave vectors were neither perfectly parallel nor perfectly perpendicular to the ordering wave vector.\\
Prior to our study we expected spectra that are strongly reminiscent of the excitations of the ferromagnetic limit. In stark contrast, we observed highly anomalous line shapes that appear to be inconsistent with any conventional scenario. For instance, shown in Fig.~\ref{figure2}(a) and (b) are typical data for $\bm{q}\parallel\bm{k}_1$ (squares in Fig.\,\ref{figure1}). Here the data are characterized by fairly broad dispersive maxima, but a naive interpretation of the data suggests an extreme form of broadening. This may be compared with data recorded for $\bm{q}\perp\bm{k}_1$ shown in Fig.\,\ref{figure2}(c) and (d) (rhombus in Fig.\,\ref{figure1}), which is essentially featureless and more like a background.  Finally, the data for the arbitrary direction shown in Fig.\,\ref{figure2}(e) and (f) display yet another type of characteristic, namely distinct maxima suggesting a well defined intense mode on a large background of scattering.\\
The remarkable variety of seemingly anomalous spectra we observe experimentally certainly does not look reminiscent of ferromagnetic spin waves at all. In a first attempt we tried to fit each scan individually with one or several Gaussians. However, this procedure failed. Recognizing that our data was taken for wave vectors much smaller than the Fermi momentum, $q \ll k_F$, in principle, the spectra should be insensitive to most microscopic details. We have therefore set up an effective model to calculate the excitation spectra based on two physically transparent and meaningful parameters, namely the ferromagnetic spin wave stiffness $c(20\,\mathrm{K})=37$~meV\AA$^2$ derived from previous experiments \cite{Ishikawa:77,semadeni:99} and the pitch of the helical modulation $k_{\rm h}=0.035$~\AA$^{-1}$. A detailed description of the theoretical model will be given in the following section~\ref{theory}. To compare the calculated spectra with our data we folded them with the energy and momentum resolution of our measurements (which is about the best currently available) and adjusted the absolute value of the intensity. We emphasize that the scaling of the intensity of all spectra by the same value represents the only free parameter of our model.\\
Our theoretical model gives complete and precise account of \textit{all measured features and lineshapes for all momenta} as illustrated in Fig.\,\ref{figure2} for six of the 24 spectra we have measured. The agreement between the calculated spectra and the experimental data can be quantified by the remarkably small value of the reduced $\chi^2$ value of $\sim1.3$. We thereby note that we used a rather conservative estimate of the instrumental resolution, which tentatively averages out small details that may possibly exist in the data. These underestimates of the experimental resolution are quite typical and observed in many inelastic neutron scattering studies. Further, when artificially modifying the values of $c$ and $k_h$ by more than $\sim5\,\%$ the quality of the agreement as measured by $\chi^2$ deteriorates rapidly. This suggests strongly that our model sensitively captures the entire physics of all excitation spectra measured.

\begin{figure}[h!]
\centering
\includegraphics[width=.47\textwidth,clip=]{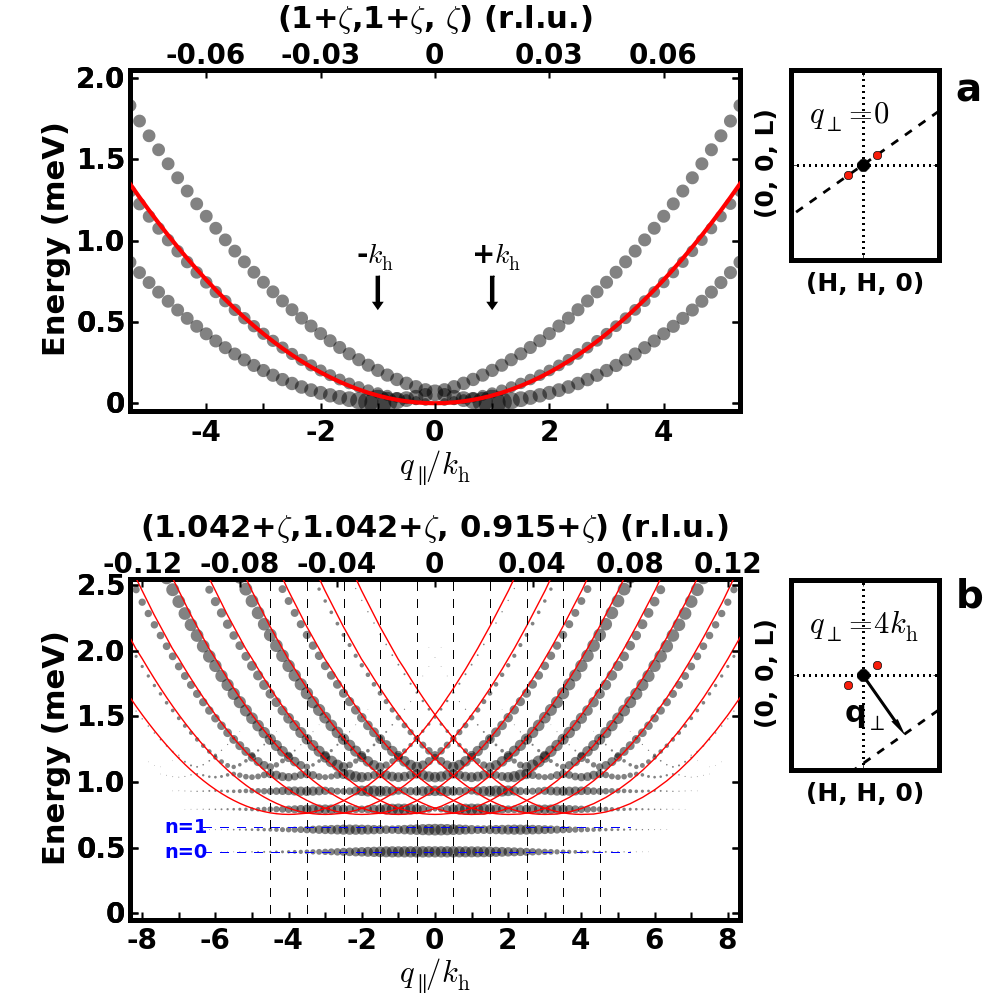}
\caption{Illustration of characteristic features of helimagnon bands. In the regime investigated, $q \gg k_{\rm h}$, a crossover to a ferromagnetic dispersion with little weight in side bands was generally expected (see text). Instead, for finite $q_\perp$ multiple bands {\em with approximately equal weight} are excited due to significant Umklapp scattering. (a) For wave vectors strictly parallel to the helix a special symmetry prohibits higher order Umklapp processes: a translation of the helix can be compensated by a rotation of all spins. As a consequence only three modes are excited, two modes with minima at $\pm \vec{k}_{\rm h}$ and an additional mode centered directly at the position of the nuclear Bragg peak with zero momentum but vanishing intensity for $\vec{q} \to 0$. For comparison the dispersion of a ferromagnetic mode $\w=cq^2$ is given (red solid line). (b) The strong Umklapp scattering for finite perpendicular momentum,  $q_\perp=4 k_h$, stops the motion of spin excitations with small $q_\|$ leading to flat bands  (see text and Eq.~(\ref{bands})). In both panels the spectral weight of the corresponding modes is proportional to the area of the points  where we use a maximal size for better visibility. For the calculation of the dispersion we used $c(20\,\mathrm{K})=37$~meV\AA$^2$ and $k_{\rm h}=0.035$~\AA$^{-1}$. For clarity only a single domain is shown,  namely $\vec{k}_1$. }\label{figure4}
\end{figure}
\begin{figure}[h!]
\centering
\includegraphics[width=.47\textwidth,clip=]{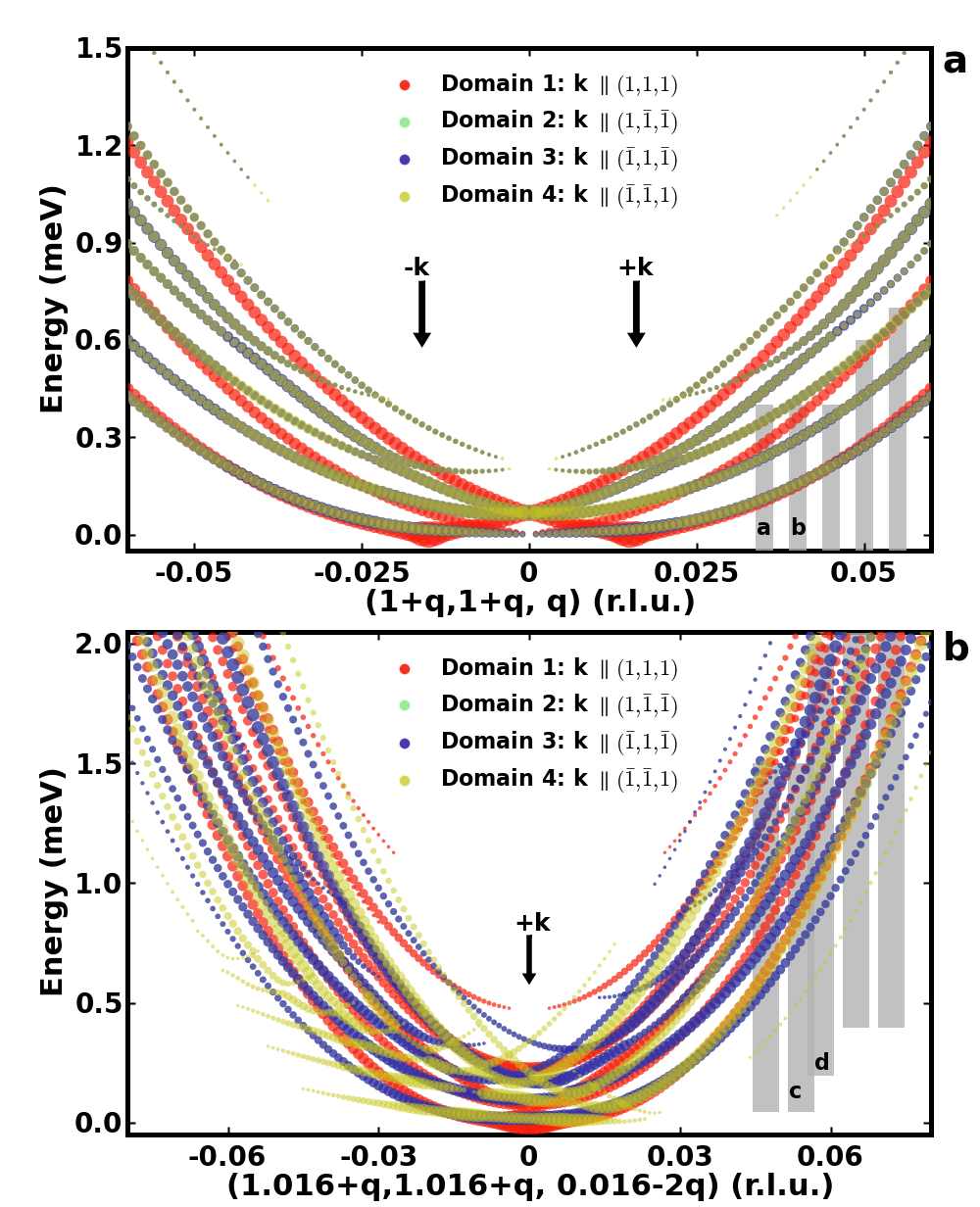}
\caption{Dispersion curves for various directions as calculated in the model described in the text.
Contributions from different domains are color coded as shown in Fig.\,\ref{figure1}. The area of each point is proportional to the weight of the corresponding peak in neutron scattering (only for the most intense peaks close to the reciprocal lattice vectors a fixed point size was used for better visibility). The complex lineshapes observed experimentally (cf. Fig.~\ref{figure2}) result from the large number of excited bands. The gray shaded bars show the direction and parameter range of the constant $Q$ scans recorded in our measurements, where the width in $Q$ of the gray area represents the resolution in $Q$. Data for the gray bars labeled a through d are shown in Fig.\,\ref{figure2}.
(a) Dispersion for the trajectory with $\vec{q}\parallel\vec{k}_1$. (b) Dispersion for the trajectory with
$\vec{q}\perp\vec{k}_1$. } \label{figure3}
\end{figure}
\section{Theory}\label{theory}
As a starting point of a detailed account of our theoretical description we emphasize again, that the weakness of spin-orbit coupling in MnSi leads to well-separated energy and length scales. The underlying physics is based on the properties of a ferromagnet. At energies below $\sim3$\,meV, where the spin-flip Stoner continuum sets in \cite{Ishikawa:77}, and for corresponding small momenta, $q<0.3 \, {\rm \AA}^{-1}$, the low-energy excitations of the ferromagnet in the absence of spin-orbit interactions  are described with high precision by a simple quadratic dispersion $E^{\rm FM}_{\vec{q}}=c q^2$. At the small wave vectors relevant for our experiment the damping of those modes may be neglected.\\
Accordingly, the starting point of our theoretical description is a rotationally invariant  non-linear $\sigma$ model (including the appropriate Berry-phase term) to describe the low-energy excitations of this ferromagnet. Following Ref.~\onlinecite{bak:80,nakanishi:80}, we add to this model the leading spin-orbit coupling effect, which is given by the rotationally invariant Dzyaloshinskii-Moriya (DM) interaction $S_{DM}= g \int \vec{\Phi} \cdot (\vec{\nabla} \times \vec{\Phi})$. Being linear in momentum, the DM interactions necessarily lead to an instability of the ferromagnet and stabilize a chiral helix of the form
\begin{equation} \label{helix} \vec{\Phi}_h(\vec{x})=\hat{\vec{n}}_1 \cos(\vec{k}_{\rm h} \cdot \vec{x})+\hat{\vec{n}}_2 \sin(\vec{k}_{\rm h} \cdot \vec{x})
\end{equation}
where the unit vectors $\hat{\vec{n}}_1$ and $\hat{\vec{n}}_2$ are perpendicular to each other and to the ordering vector $\vec{k}_{\rm h}$. In our MnSi crystals the sign of $g$ is such that $\hat{\vec{n}}_1$,
$\hat{\vec{n}}_2$ and $\vec{k}_{\rm h}$ define a left-handed coordinate system. Corrections to (\ref{helix}) are small as they arise only from tiny forth order terms in the weak spin-orbit coupling neglected in our calculation. In a ferromagnetic state these corrections lead to the formation of a gap. 
However, due to the spontanteously broken translation invariance and the Goldstone theorem there cannot be a gap for an incommensurate helimagnet. The only effect of the higher-order spin-orbit coupling terms is a modification of the spin wave spectrum for momenta much smaller than $k_p \ll \vert\vec{k}_{\rm h}\vert$, i.e., far below the experimental resolution available to date \cite{belitz:06,commentM}. \\
To describe the spin-excitations of the helix we expand $\vec\Phi$ around the mean field, $\vec\Phi=\vec{\Phi}_h+\delta \vec{\Phi}$. As discussed below the physical situation becomes transparent by using a comoving coordinate system, where the order parameter at point $\vec{x}$ is rotated by the angle $\vec{k}_{\rm h} \cdot \vec{x}$ around the $\vec{k}_{\rm h}$ axis using the space-dependent rotation matrix $R$ with $\vec{\Phi}'=R \vec{\Phi}$. In the comoving coordinate system the helix is mapped on a ferromagnetic solution $\vec{\Phi}_h'=\hat{\vec{n}}_1$. Using length and energy units such that $|\vec{k}_{\rm h}|=1$ and $E^{\rm FM}_{\vec{k}_{\rm h}}=E_{k_{\rm h}}=c k_{\rm h}^2=1$, all free parameters are fixed.\\
The long wavelength of the helix with the crystallographic lattice (which breaks translation invariance), implies a tiny magnetic  Brillouin zone and the formation of an abundance of bands. For the further discussion it is helpful to split the momentum $\tilde{q}_\|$ parallel to $\vec{k}_{\rm h}$ into $\tilde{q}_\|=n k_{\rm h}+q_\|$ introducing a band index $n$ while using $-k_{\rm h}/2<q_\|\le k_{\rm h}/2$. Denoting by $\vec{q}$ both $q_\|$ and $\vec{q}_\perp$, the Gaussian fluctuations around the mean field are described by the action
\begin{equation}
S=\frac{1}{2} \sum_{n,n',\vec{q},i \omega_n} \delta \vec{\Phi'}^*_{n\vec{q}}(\omega_n) \, M_{nn'}(\vec{q},i
\omega_n) \, \delta \vec{\Phi'}_{n'\vec{q}}(\omega_n)
\end{equation}
where each entry of $M_{nn'}$ is a two by two matrix describing the fluctuations in two spin-directions
perpendicular to the mean field $\vec{\Phi}_h$. For a ferromagnet one needs only a single matrix
\begin{eqnarray}
M^{\rm FM}(\vec{q},\w)=\left(\begin{array}{cc} q^2 & i \omega \\ - i \omega & q^2. \end{array} \right)
\end{eqnarray}
From the condition ${\rm det} M=0$ one obtains the well-known dispersion, $\w=q^2$, of a ferromagnet.\\
The fluctuations around the helimagnetic state are in contrast described by an infinite dimensional matrix, where the DM interaction yields two changes with respect to the ferromagnetic state. First, one obtains an additional ``$1$'' in the diagonal entries $M_{nn}(\vec{q},\w)$, which are given by:
\begin{eqnarray}\label{md}
M_{nn}(\vec{q},\w)=\left(\begin{array}{cc} q^2 & i \omega \\ - i \omega &
1+q^2 \end{array} \right)
\end{eqnarray}
where $q^2=q_\perp^2+(q_\|+n)^2$. Second, one obtains non-vanishing off-diagonal matrix elements given by:
\begin{eqnarray}\label{mod}
M_{n,n\pm 1}(\vec{q},\w)=\left(\begin{array}{cc} 0 & -i q_\perp^\pm \\ i q_\perp^\pm & 0 \end{array} \right)
\end{eqnarray}
with $q_\perp^\pm = q_x \pm i q_y$ for a helix oriented in $z$ direction. The simple matrix defined by Eqns.~(\ref{md},\ref{mod}) leads to the physics explained below and describes the neutron scattering experiments with high precision and no other parameters besides the (measured) pitch of the helix, the measured ferromagnetic spin wave stiffness and the overall intensity as the only free parameter.

\section{Discussion}

To compare the calculated spectra with the neutron scattering results, one determines the momentum-diagonal $3 \times 3$ susceptibility matrix $\chi_{\bf q}(\omega) \propto \langle \delta \vec{\Phi} \,\, \delta \vec{\Phi}\rangle_{\vec{q}\omega}$ by inverting first the matrix $M$ and undoing the basis change due to the rotation $R$ defined above. The scattering cross section for neutrons with a given momentum transfer $\vec{Q}=\vec{G}+\vec{q}$ close to a reciprocal lattice vector $\vec{G}$ of the MnSi crystalline lattice is proportional to $(1+n_B(\omega)) {\rm Im}\left[ {\rm Tr}(\chi_{\vec{q}}(\omega+i 0))- \hat{\bf Q} \chi_{\vec{q}}(\omega+i 0)\hat{\bf Q} \right]$, where $n_B(\omega)=1/(e^{\hbar \w/k_B T}-1)$ is the Bose function and where we have taken into account, that the neutrons couple only to the component of $\chi$ perpendicular to $\vec{Q}$. Within the Gaussian approximation one obtains sharp modes. For a comparison with experiment these theoretical results are convoluted by the experimental resolution.\\
We first consider the excitations with vanishing momentum perpendicular to the propagation vector, $\vec{q}_\perp=0$, predicted in this model. In this limit all off-diagonal matrices vanish and one obtains a single mode with the dispersion relation $\w=c q_\| \sqrt{k_{\rm h}^2+q_\|^2}$, well known from the physics of Bose condensed atoms.  It describes the crossover from a mode with linear dispersion \cite{belitz:06,maleyev:06} at small momentum to a ferromagnetic mode with quadratic dispersion. This crossover is also manifest in the eigenmodes as illustrated in Fig.\,\ref{figure1}. In the ferromagnetic limit $k_c\gg q_\|\gg k_{\rm h}$ the magnetization displays the typical precession (Fig.~\ref{figure1}b), while it oscillates only perpendicular to $q_\|$ for $q_\|\ll k_{\rm h}$ (Fig.~\ref{figure1}c).\\
By transforming back to the physical coordinate system, one realizes that generically three copies of this mode can be observed in neutron scattering with unpolarized neutrons as shown in Fig.~\ref{figure4}(a). Besides the two copies with minima at $\pm \vec{k}_{\rm h}$ one also obtains generically a mode directly at the Bragg spot with zero momentum. However, in contrast to the dispersion of a ferromagnetic mode of identical spin stiffness $c$ shown in red, in the helimagnetic case the intensity vanishes for $\vec{q} \to 0$. The fact that only three modes rather than a large number of bands can be observed, originates in the special symmetry of the helical state: a translation of the helix can be compensated by an appropriate rotation of the spins.  Theoretically this property has actually long been known \cite{nagamiya}. Only higher-order spin-orbit coupling terms which break this symmetry may lead to the excitation of further modes, however, with a tiny weight.\\
The more important second example, which reveals an entirely unexpected property, concerns spectra for finite perpendicular momentum as shown in Fig.~\ref{figure4}(b). Here the off-diagonal terms (\ref{mod}) lead to a mixing of the modes. For vanishing $q_\|$ and $q_\perp \ll k_{\rm h}$ one obtains \cite{belitz:06,maleyev:06} $E(q_\perp)=\sqrt{3/8} c q_\perp^2 $ which can, however, not be observed directly within our experimental resolution. In the limit $q_\perp \gg k_{\rm h}$ one finds that more and more modes are excited. Mapping the matrix to a harmonic oscillator on a lattice (see e.g. Ref.~\onlinecite{fischer:04}) one finds that typically of the order of $\sqrt{q_\perp/k_{\rm h}}$ bands are excited with equally spaced energies
\begin{align}\label{bands}
E_n(\vec{q})&\approx c \left(q_\perp^2-2  k_{\rm h} q_\perp+2  \sqrt{q_\perp k_{\rm h}^{3}}
(n+\frac{1}{2})\right),\nonumber\\n&=0, 1, 2, ...\sqrt{q_\perp/k_{\rm h}}.
\end{align}
Here two aspects are remarkable. First, for increasing perpendicular momentum an increasing number of bands are excited. Second, the dispersion is essentially independent of $q_\|$ for  $q_\| < \sqrt{q_\perp k_{\rm h}}$.\\
In Fig.~\ref{figure4}(b) we plot the corresponding dispersion as a function  of $q_\|$ for $q_\perp = 4 k_{\rm h}$ with $c$ and $k_{\rm h}$ given above. For clarity we consider only the domain associated with $\vec{k}_1$. The transverse momentum strongly couples the ferromagnetic branches (red solid lines), $\w(\vec{q}) = c((q_\|+m k_{\rm h})^2 + q_\perp^2)$ with $m=0,\pm 1,\pm 2 \dots$,  shifted by multiples of the reciprocal lattice vector $\vec k_{\rm h}$ of the magnetic Brillouin zone. As the DM interactions grows linearly with $q$, more an more modes are coupled by Umklapp scattering when $q_\perp$ increases. As described by Eq.~(\ref{bands}) the repeated Umklapp scattering prohibits the motion of the spin excitations parallel to $\vec k_{\rm h}$ and leads to a very flat dispersion for $q_\| < \sqrt{q_\perp k_{\rm h}}$.\\
We finally illustrate in Fig.~\ref{figure3} the full complexity of the calculated helimagnon bands as probed experimentally by the two trajectories in Fig.~\ref{figure1}. Here the area of the calculated dots is proportional to the intensity and the color coding refers to the relevant  domain. The gray shaded bars show the location and parameter range of our measurements, where the width of the gray bars represents the resolution of our measurements. Using this approach we are able to account for the entire set of very broad and extraordinarily rich spectra measured (note the very different energy dependences in Fig.~\ref{figure2}) without any further adjustments of parameters. 

\section{Conclusion}

In many magnetic materials excitation spectra like those observed here may be interpreted by invoking a host of non-universal properties. For instance, spin-orbit coupling and dipolar interactions cause magnetic anisotropies and the formation of energy gaps \cite{boeni:95}. Moreover, scattering from impurities and crystalline defects, the coupling to the particle-hole continuum in metals \cite{Ishikawa:77}, magnon-phonon \cite{Rainford:72} and magnon-magnon interactions \cite{Bayrakci:06} may lead (in combination with spin-orbit interactions) to a substantial damping of the magnetic excitations.\\ 
In summary, we report an inelastic neutron scattering study of the perhaps simplest manifestation of weak chiral interactions: the helimagnetic order in the cubic B20 compound MnSi. The excellent quantitative account we achieve establishes the existence of a new universal property of the spin excitations: the formation of a large number of dispersive bands of helimagnons. Our study is also important for the theoretical description of the skyrmion lattice in MnSi which turns out to be stabilized relative to a conical phase only by thermal fluctuations. Therefore it is important to establish that a quantitative description of the fluctuations spectrum can be obtained in this material. We expect that similar phenomena will be relevant for a broad range of materials where long wavelength or incommensurate magnetic order lead to strong Umklapp scattering and the formation of multiple bands.


\section*{Acknowledgements}

We wish to thank D. Belitz. T. R. Kirkpatrick, M. Garst, B. Binz, M. Vojta, and W. Zwerger. This research was supported in part by the National Science Foundation under Grant No. PHY05-51164 and by the SFB 608 of the DFG. The neutron scattering experiments were carried out at the continuous spallation neutron source SINQ at Paul Scherrer Institut at Villigen PSI in Switzerland and the Forschungs-Neutronenquelle Heinz Maier-Leibnitz (FRM-II) of Technische Universit\"at M\"unchen at Garching in Germany.


\begin{thebibliography}{48}
\bibitem[{com()}]{commentM}
\bibinfo{note}{In Ref. \cite{maleyev:06} it has been argued that the
  interaction of spin-waves can give rise to a gap in the spectrum. As this
  result contradicts the Goldstone theorem, we believe that it is probably not
  valid.}
\end{thebibliography}

\end{document}